\documentclass[12pt,a4paper,final]{iopart}

\usepackage{iopams}  
\usepackage[utf8]{inputenc}
\usepackage{graphicx}
\usepackage{amssymb}
\usepackage{verbatim}
\usepackage{tikz}
\usepackage{subfigure}
\usepackage{float}
\usepackage{subfig}
\usetikzlibrary{shapes.geometric, arrows}
\tikzstyle{startstop} = [rectangle, rounded corners, anchor=center, minimum width=3cm, minimum height=1cm,text centered, draw=black, fill=red!0]
\tikzstyle{arrow1} = [thick,->,>=stealth]
\tikzstyle{arrow2} = [thick,<->,>=stealth]
\tikzstyle{line} = [thick,-,>=stealth]

\begin{document}

\title[]{Discrete Cosmological Models in the Brans-Dicke Theory of Gravity}

\author{Jessie Durk and Timothy Clifton\footnote[1]{email: t.clifton@qmul.ac.uk}}
\address{School of Physics and Astronomy, Queen Mary University of London, UK}

\begin{abstract} We consider the problem of building inhomogeneous cosmological models in scalar-tensor theories of gravity. This starts by splitting the field equations of these theories into constraint and evolution equations, and then proceeds by identifying exact solutions to the constraints. We find exact, closed form expressions for geometries that correspond to the initial data for cosmological models containing regular arrays of point-like masses. These solutions extend similar methods that have recently been applied to Einstein's equations, and provides sufficient initial conditions to perform numerical integration of the evolution equations. We use our new solutions to study the effects of inhomogeneity in cosmologies governed by scalar-tensor theories of gravity, including the spatial inhomogeneity allowed in Newton's constant. Finally, we compare our solutions to their general relativistic counterparts, and investigate the effect of changing the coupling constant between the scalar and tensor degrees of freedom.

\end{abstract}

\section{Introduction}

Scalar-tensor theories of gravity are among the oldest and best studied generalisations of Einstein's theory. They were originally introduced by Jordan in 1949 \cite{jordan1, jordan2}, before being refined by Brans and Dicke in 1961 \cite{bd} and then being generalised to theories with arbitrary coupling parameters by Bergmann \cite{bergmann}, Wagoner \cite{wagoner} and Nordtvedt \cite{nordvedt}. They can be seen to contain the dimensionally reduced theories that one recovers from string theory \cite{polchinski}, as well as the canonical version of the Horndeski class of scalar-tensor theories that have recently found popularity in cosmology \cite{horndeski}. Phenomenologically, scalar-tensor theories of gravity have found application in modelling the possible variations of the constants of nature \cite{uzan1, uzan2}, as well as providing the archetypal class of theories that are used to quantify allowed deviations from Einstein's theory \cite{will}.

In this paper we study inhomogeneous cosmological solutions of the scalar-tensor theories of gravity introduced by Brans and Dicke. While much work has been performed on understanding virtually every aspect of these theories (see e.g. \cite {FM, modgrav}), it is still the case that very little is known about their cosmological solutions away from the limits of homogeneity and isotropy. To date, the only studies in this area have been limited to highly symmetric matter configurations \cite{timmodgravinhom} or theories with well chosen self-interaction potentials \cite{faraoni}. We address this deficit by studying inhomogeneous cosmological configurations that admit no global symmetries, but which allow progress to be made using exact methods. We expect the space-times that result from our investigation to shed light on the consequences of structure formation in these theories, including the degree to which Newton's constant is allowed to vary in space.

In general, the effect that large-scale structures have on the expansion of the Universe has proven to be a subject of much contemporary interest \cite{clarkson, buchert, clifton}, yet has so far only really been studied in the context of Einstein's equations. The interest in this subject arises principally due to the non-commutativity of averaging and evolution under non-linear field equations, which means the large-scale average expansion of an inhomogeneous Universe can evolve in a non-trivial way. This has potentially serious implications for the interpretation of data in the real Universe, where the effects of inhomogeneity have been suggested to have consequences for everything from the existence of dark energy \cite{rasanen} to the recent tension between local and global measurements of the Hubble constant \cite{bolejko}.

Well defined inhomogeneous cosmological models are needed to study these possibilities, and to precisely quantify any effects that arise. Various approaches have been taken to construct such models in recent years, including the application of numerical, perturbative and analytic techniques. Here we are interested in the set of models that have come to be known as `black hole lattices' \cite{durkreview}. These are based on the Lindquist-Wheeler models first proposed in Ref. \cite{lw}, and which describe a closed universe filled with Schwarzschild-like masses (or black holes) as an initial value problem. The construction of initial data in these situations can often be performed analytically, if the initial hypersurface is taken to be extrinsically flat \cite{tim1} or has constant mean curvature \cite{durklambda}, and is often sufficient to determine some of the large-scale properties of the cosmology as a whole. It also provides the basis for investigating the evolution of such a universe, using either perturbative \cite{tim2, tim3} or numerical techniques \cite{eloisa1, eloisa2, yoo2, yoo3}.

In this paper we apply the techniques developed in the study of general relativistic black hole lattices to the Brans-Dicke theory of gravity, to find exact initial data for universes that contain arrays of regularly arranged point-like masses in universes governed by these theories. This extends the results of previous studies to new theories of gravity, allowing the general relativistic results to be considered within a wider context. It also significantly extends what is currently known about inhomogeneous cosmological models in scalar-tensor theories of gravity - a field that is severely restricted by the additional complexity of the field equations. 

This paper is organised as follows: In Section \ref{sec:bd} we briefly review the Brans-Dicke theory of gravity, before deriving the relevant constraint equations for our initial data problem. In Section \ref{sec:constraints} we investigate solutions to these equations, including expressions for the proper masses and scalar charges for each of the point-like objects. Section \ref{sec:flrw} then contains a review of Friedmann cosmology in Brans-Dicke theory, and proceeds to compare the scale of our inhomogeneous models to these perfectly homogeneous and isotropic solutions. Finally, we conclude in Section \ref{sec:discussion}.

\section{The Brans-Dicke theory of gravity}
\label{sec:bd}

\subsection{Field equations}

The Brans-Dicke scalar-tensor theory of gravity requires us to introduce an additional scalar field $\phi$, as well as the metric $g_{\mu\nu}$, in the Lagrangian density:
\begin{equation}
    \mathcal{L} = \frac{1}{16\pi}\sqrt{-g}\left(\phi R - \frac{\omega}{\phi}\nabla_\mu \phi \nabla^\mu \phi \right) + \mathcal{L}_m(g_{\mu \nu}, \psi),
\end{equation}
where $\omega$ is the constant coupling parameter of the theory, and $\mathcal{L}_m(g_{\mu \nu}, \psi)$ is the Lagrangian density of the matter fields, $\psi$. The non-minimal coupling between $\phi$ and $R$ results in new gravitational phenomena, while the coupling of only $g_{\mu \nu}$ to $\psi$ ensures that the Einstein equivalence principle is maintained. 

Varying the resulting action with respect to the metric $g_{\mu\nu}$, gives the following field equations:
\begin{equation}
    \phi \, G_{\mu\nu} + \left(\Box \phi + \frac{\omega}{2 \phi}(\nabla \phi)^2 \right)g_{\mu\nu} - \nabla_\mu \nabla_\nu \phi - \frac{\omega}{\phi}\nabla_\mu \phi \nabla_\nu \phi = 8 \pi \, T_{\mu \nu},
    \label{eq:fe1}
\end{equation}
while varying with respect to the scalar field $\phi$, yields
\begin{equation}
    \Box \phi = \frac{8 \pi \, T}{(3+2 \omega)}, 
        \label{eq:fe1b}
\end{equation}
where $T_{\mu \nu}$ are the components of the energy-momentum tensor, and $T$ is its trace. The locally measured gravitational `constant' in these theories can then be shown to be given by
\begin{equation} \label{newtonG}
    G= \frac{(4+2 \omega)}{(3+2 \omega)} \frac{1}{\phi},
\end{equation}
and hence can vary in space-time whenever $\phi$ is non-constant. These equations can be seen to reduce to Einstein's theory in the limit $\omega\rightarrow \infty$, when $\phi \rightarrow$ constant, and Equation (\ref{eq:fe1}) reduces to Einstein's equations.

\subsection{Constraint equations in vacuum}

We now wish to derive the Hamiltonian and momentum constraint equations that correspond to the field equations in (\ref{eq:fe1}) and (\ref{eq:fe1b}). This is done by performing the usual $3+1$ decomposition, using the irrotational time-like unit normal $n_{\mu}$ and the projection tensor $h_{\mu \nu} = g_{\mu \nu} +n_{\mu} n_{\nu}$. All quantities can then be split into a temporal part, by contracting with $n_{\mu}$, and a spatial part, by projecting with $h_{\mu \nu}$. In particular, the Gauss-Codazzi-Mainardi equations can be used to project the Einstein tensor such that
\begin{equation}
\label{GC1}
 2 G_{\mu \nu} n^{\mu} n^{\nu} =   {}^{(3)}R + K^2 -K_{\mu \nu}K^{\mu\nu},
\end{equation}
where $K_{\mu\nu}=-h_{\mu}^{\phantom{\mu} \rho}h_{\nu}^{\phantom{\nu} \sigma} \nabla_{\rho} n_{\sigma}$ is the extrinsic curvature of the hypersurfaces orthogonal to $n^{\mu}$, $K$ is its trace, and ${}^{(3)}R$ is the Ricci curvature scalar of the space orthogonal to $n^{\mu}$. As well as this we find
\begin{equation}
\label{GC2}
  -h_{\mu}^{\,\,\nu}G_{\nu \sigma}n^{\sigma}=  D_{\nu} K^{\nu}_{\,\,\mu} - D_{\mu} K ,
\end{equation}
where $D_{\mu}$ is the torsion-free covariant derivative on the hypersurface orthogonal to $n^\mu$ that is compatible with $h_{\mu\nu}$, and which is defined such that $D_{\mu} K_{\nu \rho} = h_{\mu}^{\,\, \sigma} h_{\nu}^{\,\, \tau} h_{\rho}^{\,\, \chi} \nabla_{\sigma} K_{\tau \chi}$ (for example). For general relativity in vacuum, the left-hand sides of equations (\ref{GC1}) and (\ref{GC2}) are zero. For Brans-Dicke theory, however, this will not be true -- the left-hand side will instead be a function of the scalar field, $\phi$. 

When $T_{\mu \nu}=0$, we can use Equation (\ref{eq:fe1}) to write the Hamiltonian constraint as
\begin{eqnarray}
\label{eq:deriv1}
  \hspace{-2cm} ^{(3)} R + K^2 -K_{\mu \nu}K^{\mu \nu} 
    =  2 \frac{\Box \phi}{\phi} +  \frac{2}{\phi}n^{\mu} n^{\nu} \nabla_{\mu} \nabla_{\nu} \phi + \frac{\omega}{ \phi^2}(\nabla \phi)^2 + \frac{2\omega}{\phi^2} n^{\mu} n^{\nu} \nabla_{\mu} \phi \nabla_{\nu} \phi\, ,
\end{eqnarray}
where we have used $g_{\mu \nu}n^{\mu} n^{\nu} = -1$. The first and second terms on the right-hand side of Equation (\ref{eq:deriv1}) can then be used to write
\begin{eqnarray*}
    &\frac{2}{\phi}\left(\Box \phi + n^{\mu} n^{\nu}  \nabla_{\mu} \nabla_{\nu} \phi\right) \\
    =& \frac{2}{\phi} K n_{\mu} \nabla^{\mu} \phi + \frac{2}{\phi}\left(- K n_{\mu} \nabla^{\mu} \phi + g^{\mu\nu}\nabla_{\mu} \nabla_{\nu} \phi + n^{\mu} n^{\nu}  \nabla_{\mu} \nabla_{\nu} \phi\right)\\
    =& \frac{2}{\phi} K n_{\mu} \nabla^{\mu} \phi + \frac{2}{\phi}\left(- K n_{\mu} \nabla^{\mu} \phi + h^{\mu \nu}\nabla_{\mu} \nabla_{\nu} \phi \right)\\
    =& \frac{2}{\phi} K n_{\mu} \nabla^{\mu} \phi + \frac{2}{\phi}\left(\nabla_{\nu} (n^{\rho} n_{\mu})h_{\rho}^{\,\,\nu}\nabla^{\mu} \phi + h^{\mu\nu}\nabla_{\mu} \nabla_{\nu} \phi \right)\\
     =& \frac{2}{\phi} K n_{\mu} \nabla^{\mu} \phi + \frac{2}{\phi}\left(\nabla_{\nu} (h_{\mu}^{\,\,\rho})h_{\rho}^{\,\,\nu}\nabla^{\mu} \phi + h^{\mu \nu}\nabla_{\mu} \nabla_{\nu} \phi \right)\\
      =& \frac{2}{\phi} K n_{\mu} \nabla^{\mu} \phi + \frac{2}{\phi}\left(\nabla_{\nu}(h_{\mu}^{\,\,\rho}\nabla^{\mu} \phi)h_{\rho}^{\,\,\nu} \right)\\
     =& \frac{2}{\phi} K n_{\mu} \nabla^{\mu} \phi + \frac{2}{\phi}D_{\mu} D^{\mu} \phi\, ,
    \end{eqnarray*}
while the third and fourth terms can be written as 
\begin{eqnarray*}
      &  \frac{\omega}{\phi^2}\left( g^{\mu\nu}\nabla_{\mu} \phi \nabla_{\nu} \phi + 2 n^{\mu} n^{\nu}  \nabla_{\mu} \phi \nabla_{\nu}\phi \right) \\
      =& \frac{\omega}{\phi^2}\left( (h^{\mu\nu} - n^{\mu} n^{\nu})\nabla_{\mu} \phi \nabla_{\nu} \phi + 2n^{\mu} n^{\nu}\nabla_{\mu} \phi \nabla_{\nu}\phi  \right) \\
        =& \frac{\omega}{\phi^2}\left( h^{\mu\nu}\nabla_{\mu} \phi \nabla_{\nu} \phi + n^{\mu} n^{\nu} \nabla_{\mu} \phi \nabla_{\nu} \phi \right) \\
        =& \frac{\omega}{\phi^2}\left( D_{\mu} \phi D^{\mu} \phi + n^{\mu} \nabla_{\mu} \phi \, n^{\nu} \nabla_{\nu}\phi \right)\, .
    \end{eqnarray*}
Combining these results we then have that the Hamiltonian constraint can be written as
\begin{equation} \label{hamfinal}
    ^{(3)} R + K^2 -K_{\mu\nu}K^{\mu\nu} - 2 K \frac{\dot{\phi}}{\phi} - \frac{2}{\phi}D^2 \phi -\frac{\omega}{\phi^2} \dot{\phi}^2 -\frac{\omega}{\phi^2}D_{\mu} \phi \, D^{\mu} \phi = 0 \, ,
\end{equation}
where $\dot{\,}=n^{\mu} \nabla_{\mu}$ and $D^2 = D_{\mu} D^{\mu}$. Similarly, for the momentum constraint we have 
\begin{eqnarray} 
  &D_{\mu} K^{\mu}_{\,\,\nu} - D_{\nu} K  \\\nonumber  
    =&   \frac{\Box \phi}{\phi}h_{\nu}^{\,\,\mu} n_{\mu} + \frac{\omega}{2}\frac{(\nabla \phi)^2}{\phi^2}h_{\nu}^{\,\,\mu} n_{\mu} - \frac{\omega}{\phi^2} h_{\nu}^{\,\,\mu} n^{\rho} \nabla_{\mu} \phi \, \nabla_{\rho} \phi  - \frac{1}{\phi} h_{\nu}^{\,\,\mu} n^{\rho} \nabla_{\mu} \nabla_{\rho} \phi  \, .
    \end{eqnarray}  
The first and second terms on the right-hand side of this equation can immediately be seen to vanish, as $h_{\nu}^{\,\,\mu}n_{\mu} = 0$. The third term, on the other hand, is simply $- \omega \dot{\phi} D_{\nu} \phi/\phi^2$. Finally, the last term can be written as 
\begin{eqnarray} \nonumber
     -\frac{1}{\phi}h_{\nu}^{\,\,\mu}n^{\rho} \nabla_{\mu} \nabla_{\rho} \phi &= -\frac{1}{\phi}h_{\nu}^{\,\,\mu} \nabla_{\mu} (n^{\rho} \nabla_{\rho} \phi) +\frac{1}{\phi}h_{\nu}^{\,\,\mu} (\nabla_{\mu} n^{\rho}) \nabla_{\rho} \phi\\
     &= -\frac{1}{\phi}D_{\nu} \dot{\phi} -\frac{1}{\phi}K_{\nu}^{\,\,\rho} \nabla_{\rho} \phi \, . \nonumber
\end{eqnarray}
Combining these results we have that the momentum constraint can be written as
\begin{equation} \label{momfinal}
    D_{\mu} K^{\mu}_{\,\,\nu} - D_{\nu} K + \frac{1}{\phi}K_{\nu}^{\,\,\mu} D_{\mu} \phi + \frac{1}{\phi}D_{\nu} \dot{\phi} +  \frac{\omega}{\phi^2} \dot{\phi} D_{\nu} \phi = 0 \, .
\end{equation}
Equations (\ref{hamfinal}) and (\ref{momfinal}) are the final version of the Hamiltonian and momentum constraint equations we wish to use, and can be seen to be consistent with other similar results derived in the literature \cite{constraints}. 

Finally, we wish to write the scalar field equation (\ref{eq:fe1b}) as a set of constraint and evolution equations. This is most conveniently done by introducing the new variables $\pi \equiv \dot{\phi}$ and ${\psi}_{\mu}=D_{\mu} \phi$. The set of evolution equations for $\phi$, ${\pi}$ and ${\psi}_{\mu}$ are then given in vacuum by
\begin{eqnarray*}
\dot{\phi} &= {\pi} \\
\dot{{\pi}} &= D_{\mu} \psi^{\mu} + K \pi + \dot{n}^{\mu} \psi_{\mu} \\
\dot{\psi}_{\mu} &= D_{\mu} {\pi} + \dot{n}_{\mu} \pi + n_{\mu} \dot{n}^{\nu} {\psi}_{\nu} + K_{\mu}^{\phantom{\mu} \nu} {\psi}_{\nu} 
\end{eqnarray*}
with the only constraint being
\begin{equation*}
    \psi_{\mu}-D_{\mu} \phi =0 \, .
\end{equation*}
This last equation is, of course, just the definition of the variable $\psi_{\mu}$, and must therefore be satisfied identically. We note that these equations are the same as those considered in Ref. \cite{carr}, for a minimally coupled scalar field in Einstein's theory{\footnote{Except for a missing term $+K_{\mu}^{\phantom{\mu} \nu} {\psi}_{\nu}$, on the right-hand side of Eq. (13) of that paper.}}. 

The only equations that need to be satisfied, in order to fully specify the initial data of a vacuum space-time in this theory, are therefore just (\ref{hamfinal}) and (\ref{momfinal}). In the next section of this paper we will solve these equations in order to find initial data for a universe filled with point-like masses.

\section{Initial data}
\label{sec:constraints}

\subsection{Time-symmetric initial data}

In order to simplify the constraint equations we can choose the extrinsic curvature to vanish, such that $K_{\mu\nu} = 0$. A hypersurface that satisfies this condition is time-symmetric, and in a cosmological context corresponds to a maximum of expansion. It also provides an analogous situation to the general relativistic studies that have already been performed for this situation \cite{tim1}. In this case the Hamiltonian and momentum constraint equations then become

\begin{equation} \label{hamc}
    ^{(3)} R - \frac{2}{\phi}D^2 \phi -\frac{\omega}{\phi^2}\dot{\phi}^2 -\frac{\omega}{\phi^2}D_a \phi D^a \phi = 0 \, ,
\end{equation}
and
\begin{equation} \label{momc}
    \frac{1}{\phi}D_a \dot{\phi} + \frac{\omega}{\phi^2} \dot{\phi} D_a \phi = 0 \, ,
\end{equation}
and where we are now using Latin indices to denote coordinates on the 3-dimensional initial hypersurface, such that $D_a$ is a covariant derivative with respect to the metric $h_{ab}$ of this space.

At this point we can see that Equation (\ref{momc}) is satisfied if either $\dot{\phi}=0$ or $\dot{\phi} \propto \phi^{-\omega}$. The former of these corresponds to a scalar field that is also time-symmetric at the initial hypersurface. The latter case is not time-symmetric, and offers a potentially interesting scenario to study, but in this case we are unable to find solutions to the corresponding Hamiltonian equation. We therefore restrict our attention to the $\dot{\phi}=0$ case, for which the Hamiltonian constraint (\ref{hamc}) becomes
\begin{equation}
    ^{(3)} R = (\omega + 2)\tilde{\psi}_a\tilde{\psi}^a + 2 D_a \tilde{\psi}^a\, ,
    \label{eq:hcI}
\end{equation}
where we have defined $\tilde{\psi}_a \equiv \psi_a/\phi = D_a \phi/\phi$. This single equation is a profound simplification of the initial system of constraint equations, but it is still a non-linear differential equation for the variable $\tilde{\psi}$ in terms of the 3-curvature $^{(3)} R$. We will now show that through a change of variables we can express this as a set of linear equations, which therefore admit solutions that can be linearly superposed.

Let us now suppose that the geometry of initial hypersurface can be written as
\begin{equation}
    ds^2 = \Omega^4(r,\theta,\varphi) \, d\bar{s}^2_3
    \label{eq:hypersurface}
\end{equation}
where $d\bar{s}^2_3 = dr^2 + \sin^2{r}(d\theta^2+\sin^2{\theta}d\varphi^2)$ is the line-element of a hypersphere, and $r$, $\theta$ and $\varphi$ are hyperspherical polar coordinates. A positive spatial curvature of this kind is required for the posited maximum of expansion, and in a conformal geometry of this type the 3-curvature $^{(3)}R$ becomes $\Omega^{-4} \,^{(3)}\bar R - 8\Omega^{-5} \bar{D}^2 \Omega$, where $^{(3)}\bar R$ is the 3-curvature of the conformal hypersurface (which equals six for a 3-sphere), and where $\bar{D}^2$ is the Laplacian on the conformal hypersurface (see equations (3.5)-(3.11) in Ref. \cite{baum}).

The change of variables we wish to perform is then given by
\begin{eqnarray}
    \Omega &= \chi^{a} \sigma^{1-a} \qquad {\rm and} \qquad \phi = \chi^{s} \sigma^{-s} \, ,
    \label{eq:cov}
\end{eqnarray}
where $a$ is a constant, $s=(1-2a \pm \tau)/(2+\omega)$ and $\tau = \sqrt{1 + 4 a(1 - a)  (3 + 2 \omega)}$. In this case, the constraint equation (\ref{eq:hcI}) is satisfied by any solutions of the following two linear equations:
\begin{eqnarray} \label{helm1}
    \bar{D}^2\sigma &= \kappa_1\, \sigma\\ \label{helm2}
    \bar{D}^2\chi &= \kappa_2 \,\chi \, ,
\end{eqnarray}
where $\bar{D}^2$ is the Laplacian operator on the conformal hypersphere described by $d\bar{s}^2$ and $\kappa_1$ and $\kappa_2$ are constants. If we choose $s=(1-2a - \tau)/(2+\omega)$, then $\kappa_1$ satisfies
\begin{equation}
    \kappa_1 = \frac{3(2+\omega)-(1+2a(3+2\omega)-\tau)\kappa_2}{(7+4\omega-2a(3+2\omega)+\tau)} \,.
\end{equation}
If one were to choose $s=(1-2a + \tau)/(2+\omega)$, then the sign of $\tau$ would need to also be changed in this expression. However, in what follows we will use the first choice of $s$ (this will be explained when we compare our solution to known exact solutions).
 
Equations (\ref{helm1}) and (\ref{helm2}) are both Helmholtz equations, which have the following smooth solutions \cite{carr}:
\begin{eqnarray}\label{sol1}
     \sigma(r, \theta, \varphi) &= \sum_i^N \alpha_i \, \frac{\sin{\{\sqrt{1-\kappa_1}(\pi-r_i)\}}}{\sin{\{\sqrt{1-\kappa_1}\,\pi\}}\sin{\{r_i\}}},\\ \label{sol2}
     \chi(r, \theta, \varphi) &= \sum_i^N \gamma_i \, \frac{\sin{\{\sqrt{1-\kappa_2}(\pi-r_i)\}}}{\sin{\{\sqrt{1-\kappa_2}\,\pi\}}\sin{\{r_i\}}},
\end{eqnarray}
where $\{\alpha_i\}$ and $\{\gamma_i\}$ are two sets of constants. Each of the terms in each of these two sums can be seen to diverge at $r_i=0$, and remain smooth and single valued everywhere else. Both $\sigma$ and $\chi$ therefore contain $N$ poles, which we take to be located at $N$ distinct locations on the conformal hypersphere. The meaning of $r_i$, as used in each of the different terms in these two equations, should therefore be taken to mean the value of the $r$ coordinate after rotating coordinates so that the pole for that particular term appears at $r=0$. In this sense, we are using a different set of hyperspherical polar coordinates for each term, so that we can write every term in the same form.

\subsection{Comparison with the Brans solution}

The solutions given in Equations (\ref{sol1}) and (\ref{sol2}) contain $2N+3$ free parameters: $\alpha_i, \gamma_i, \omega, \kappa_2$ and $a$. At this point it is instructive to compare our solution with the spherically symmetric, vacuum Brans solution, in order to understand these degrees of freedom. The line-element for the Brans solution is given by \cite{modgrav}
\begin{equation}
    ds^2 = -e^{2\alpha_0}\left(\frac{1-\frac{B}{r}}{1+\frac{B}{r}}\right)^{\frac{2}{\lambda}}dt^2 + e^{2 \beta_0}\left(1 + \frac{B}{r} \right)^4 \left(\frac{1-\frac{B}{r}}{1+\frac{B}{r}} \right)^{\frac{2(\lambda - c - 1)
    }{\lambda}} d\bar{s}^2,
    \label{eq:bd}
\end{equation}
where $d\bar{s}^2 = dr^2 + r^2(d\theta^2+\sin^2{\theta}d\varphi^2)$, $\lambda^2 \equiv (c+1)^2 - c(1-\omega\, c/2)$ and $c, B, \alpha_0$ and $\beta_0$ are constants. This solution also has a scalar field $\phi$ which can be written as 
\begin{equation}
    \phi = \phi_0 \left( \frac{1-\frac{B}{r}}{1+\frac{B}{r}}\right)^{\frac{c}{\lambda}},
\end{equation}
where $\phi_0$ is another constant. By comparing this solution with Equations (\ref{eq:hypersurface}) and (\ref{eq:cov}) and requiring that $s=(1-2a - \tau)/(2+\omega)$, the following identification can be made:
\begin{equation}
    s = \frac{c}{\lambda} \qquad {\rm and} \qquad a = \frac{\lambda-c-1}{2\lambda} \, .
\end{equation}
For the choice of $s=(1-2a + \tau)/(2+\omega)$, our solution satisfies the constraint equations if we identify $s$ with $-c/\lambda$ and $a$ with $1 - (\lambda - c - 1)/\lambda$. This shows that the choice in the sign of $\tau$ in the parameter $s$ is degenerate with the identification of $\chi$ and $\sigma$ with either $\left(1-\frac{B}{r} \right)$ or $\left( 1+\frac{b}{r} \right)$.

Now it is known that if $c = -{1}/({2+\omega})$, then Equation (\ref{eq:bd}) reduces to the Schwarzschild solution as $\omega \rightarrow \infty$ \cite{bd}. Making this choice for $c$ then gives us
\begin{equation}
    \hspace{-1cm} s = -\frac{\sqrt{2}}{\sqrt{2+\omega}\sqrt{3+2\omega}} \qquad {\rm and} \qquad a = \frac{1}{2} - \frac{1+\omega}{\sqrt{2}\sqrt{2+\omega}\sqrt{3+2\omega}} \, ,
\end{equation}
as well as
\begin{equation}
    \kappa_1 = \frac{3-2\kappa_2+\kappa_2\sqrt{\frac{6+4\omega}{2+\omega}}}{2+\sqrt{\frac{6+4\omega}{2+\omega}}} \, ,
    \label{eq:k1k2def}
\end{equation}
which can be seen to become $s=a=0$ and $\kappa_1=3/4$ in the limit $\omega \rightarrow \infty$. This  reduces the number of free parameters in our solutions to $2 (N+1)$: $\alpha_i, \gamma_i, \omega$ and $\kappa_2$. We will further investigate the meaning of these remaining degrees of freedom in what follows.

\subsection{Proper mass}

In order to determine the proper mass of each of the point-like objects in our solution, we need to view them from infinity in the asymptotically flat region on the far side of the Einstein-Rosen bridge. This means taking the limit $r_i \to 0$, which gives 
\begin{equation} \label{massds}
    ds^2 \to \left(\frac{\gamma_i}{r_i} + B_i \right)^{4a}\left(\frac{\alpha_i}{r_i} + A_i \right)^{4-4a} d\bar{s}^2,
\end{equation}
where 
\begin{eqnarray}
\label{eq:bigai}
    A_i &= - \frac{\alpha_i \sqrt{1-\kappa_1}}{\tan{\{\sqrt{1-\kappa_1}\pi\}}} + \sum_{j \neq i}\alpha_j \frac{\sin{\{\sqrt{1-\kappa_1}(\pi-r_{ij})\}}}{\sin{\{\sqrt{1-\kappa_1}\pi\}}\sin{\{r_{ij}\}}} \, ,\\
     B_i &= - \frac{\gamma_i \sqrt{1-\kappa_2}}{\tan{\{\sqrt{1-\kappa_2}\pi\}}} + \sum_{j \neq i}\gamma_j \frac{\sin{\{\sqrt{1-\kappa_2}(\pi-r_{ij})\}}}{\sin{\{\sqrt{1-\kappa_2}\pi\}}\sin{\{r_{ij}\}}} \, ,
     \label{eq:bigbi}
\end{eqnarray}
where $r_{ij}$ is the coordinate distance between points $i$ and $j$ (after rotating so that mass $i$ appears at $r = 0$). We have also used the fact that in the limit $r_i \to 0$, then $d \bar{s}^2_3 \to d\bar{s}^2$ as $\sin^2{r} \to r^2$. If we now define a new coordinate $r'_i \equiv \alpha_i^{\,2-2a}\gamma_i^{\,2a}/r_i $, it can immediately be seen that in the limit $r_i \to 0$ we have $r_i' \to \infty$. Inserting this into Equation (\ref{massds}) gives
\begin{eqnarray} \label{eq:lattice-mass}
    ds^2 
    &\rightarrow \left(1+4\frac{(1-a)\alpha_i^{\,1-2a}\gamma_i^{\,2a} A_i + a \alpha_i^{\,2-2a}\gamma_i^{\,2a-1} B_i}{r'_i}\right) d\bar{s}^{\,\prime 2} \, ,
\end{eqnarray}
where $d\bar{s}^{\,\prime 2} = dr'_i{}^2 +r'_i{}^2 d\Omega^2$. Similarly, in the limit $r \to \infty$ the static, spherically symmetric Brans solution in Equation (\ref{eq:bd}) becomes 
\begin{eqnarray}
     ds^2 
     &\rightarrow e^{2 \beta_0}\left(1 + 4\frac{(c+1)}{\lambda}\frac{B}{r} \right) d\bar{s}^2.
     \label{eq:bd-mass}
\end{eqnarray}
which, up to an overall constant rescaling of units, can be compared to Equation (\ref{eq:lattice-mass}) to give
$
  {B(c+1)}/{\lambda} = (1-a)\alpha_i^{\,1-2a}\gamma_i^{\,2a} A_i + a \alpha_i^{\,2-2a}\gamma_i^{\,2a-1} B_i
$.
We now recall that the parameter $B$ in the Brans solution is related to its mass $m$ by $ B = {m \lambda}/{2}$ \cite{bd}. Recalling $c=-1/(2+\omega)$, we can now read off that
\begin{equation}
    m_i  = 2\left(\frac{2+\omega}{1+\omega}\right)\left(\frac{\gamma_i}{\alpha_i}\right)^{2a-1} \left( (1-a)\gamma_i A_i + a \alpha_i B_i \right) \, .
    \label{eq:propermass}
\end{equation}
We take this to be the proper mass of each of the point masses in our solution.

\subsection{Scalar charge}

As well as mass, we can also derive an expression for the scalar charge, $q_i$, of each of the objects in our solution. For this we will define the scalar charge to be given by
\begin{equation}
    q_i \equiv \frac{1}{4\pi} \int \phi_{,a}n^a dA \, ,
\end{equation}
where $n^a$ is the unit inward pointing normal and $dA$ is an area element as $r\rightarrow 0$, such that $n^a = (-\sigma^{-2+2a}\chi^{-2a},0,0)$ and $dA = \sigma^{4-4a}\chi^{4a}r^2 \sin{\theta}\, d\theta\,  d\varphi$. Just as for the proper mass, we evaluate $\sigma$ and $\chi$ in the asymptotic limit $r \to 0$. This gives an expression for the scalar charge of the $i$th mass as
\begin{equation}
    q_i = s \left(\frac{\gamma_i}{\alpha_i}\right)^{s+2a-1} \left( \gamma_i A_i - \alpha_i B_i \right) \, ,
    \label{eq:scalarcharge}
\end{equation}
which has a pleasing symmetry with the expression for the proper masses given in Equation (\ref{eq:propermass}). It is straightforward to verify that in the limit $\omega \to \infty$, we recover $q_i \to 0$, as expected. These results show that the proper mass, $m_i$, and scalar charge, $q_i$, of each mass are directly related to the values of the parameters $\alpha_i$ and $\gamma_i$, and that by specifying that value of $m_i$ and $q_i$ for each of our points we are essentially setting the values of $\alpha_i$ and $\gamma_i$. This leaves only the values of $\omega$ and $\kappa_2$ as the remaining two degrees of freedom. The former of these corresponds to a choice of the gravitational theory being considered, as it appears as a coupling constant in the generating Lagrangian. We interpret the latter as corresponding to the amount of scalar field in the background cosmology, as explained below.

\subsection{Background scalar field}

Figures 1 - 4 depict the conformal factor $\Omega$ and scalar field $\phi$ for different choices of the parameter $\kappa_2$. In each of these diagrams we have set $\omega = \alpha = \gamma = 1$, and taken a surface at $r=\pi/2$ in a lattice of eight point-like masses at the following coordinate positions $(r,\theta,\varphi)$:

\vspace{-0.5cm}
\begin{table}[h!]
\begin{center}
\begin{tabular}{cccc}
$\left( 0,\frac{\pi}{2},\frac{\pi}{2} \right)$ & $\left( \frac{\pi}{2}, 0, \frac{\pi}{2} \right)$ & $\left(\frac{\pi}{2}, \frac{\pi}{2}, 0\right)$ & $\left(\frac{\pi}{2}, \frac{\pi}{2}, \frac{\pi}{2}\right)$ \\
$\left( \pi, \frac{\pi}{2} , \frac{\pi}{2} \right)$ & $\left( \frac{\pi}{2}, \pi, \frac{\pi}{2}\right)$ & $\left(\frac{\pi}{2}, \frac{\pi}{2}, \pi\right)$ & $\left(\frac{\pi}{2}, \frac{\pi}{2}, \frac{3\pi}{2}\right)$
\end{tabular}
\end{center}
\end{table}
\vspace{-0.5cm}

\noindent
Such an arrangement of points are all equally spaced from their nearest neighbours, and hence constitute a regular lattice on the hypersphere. The slice taken through this configuration in Figures 1 - 4 is a great sphere, and is chosen so that six of the eight points are positioned within that sphere, shown by the tubes in the figures representing the conformal factor $\Omega$. In each of the figures, the distance of the surface from the centre is the value of the field ($\Omega$ or $\phi$) at that point, whilst the angular positions of the surfaces correspond to individual points on the great sphere $r = \pi/2$.

\begin{figure}[t]
    \centering
    \begin{minipage}{0.3\textwidth}
        \centering
        \includegraphics[width=1\textwidth]{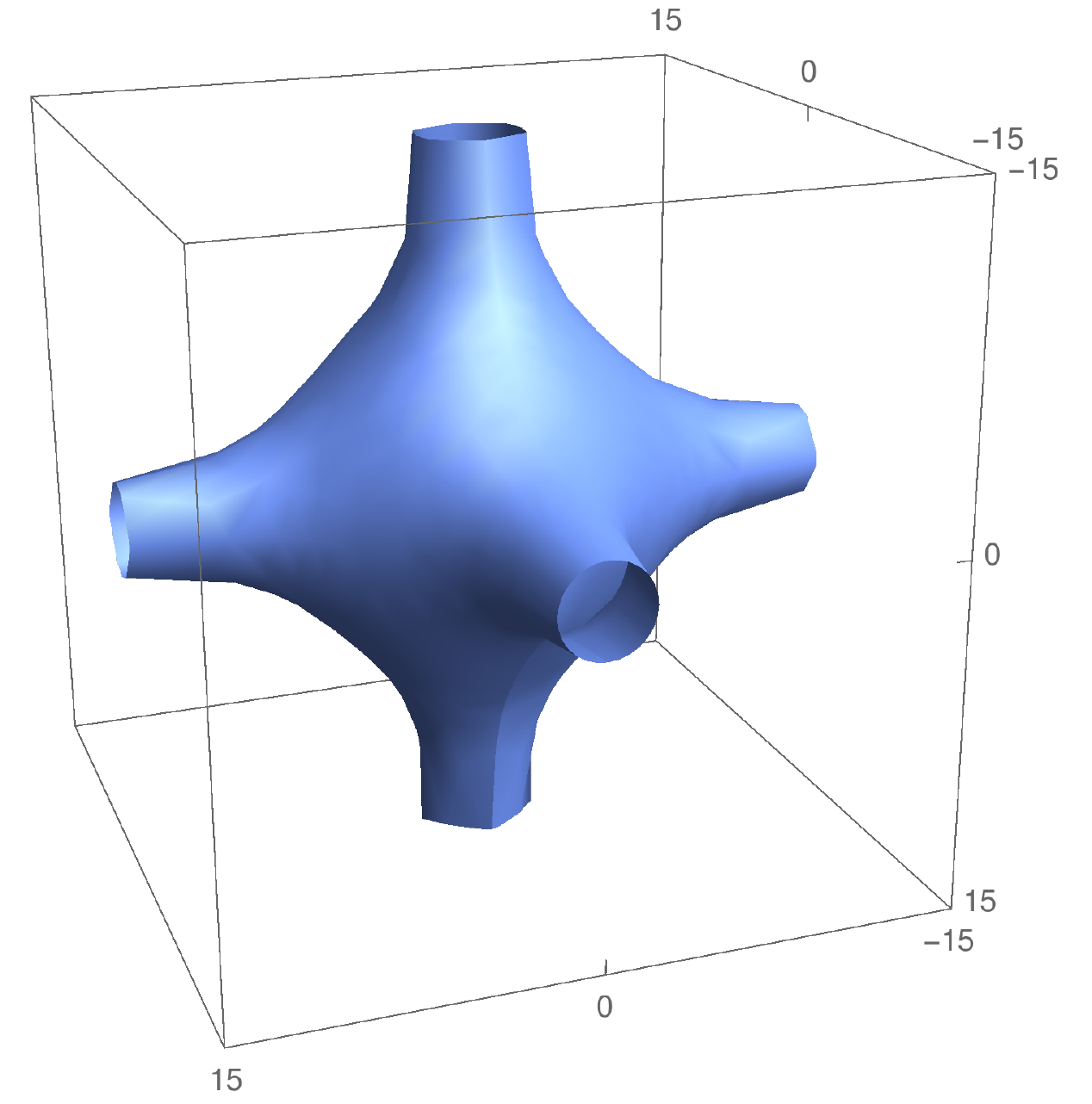} 
    \end{minipage}
    \begin{minipage}{0.3\textwidth}
        \centering
        \includegraphics[width=1\textwidth]{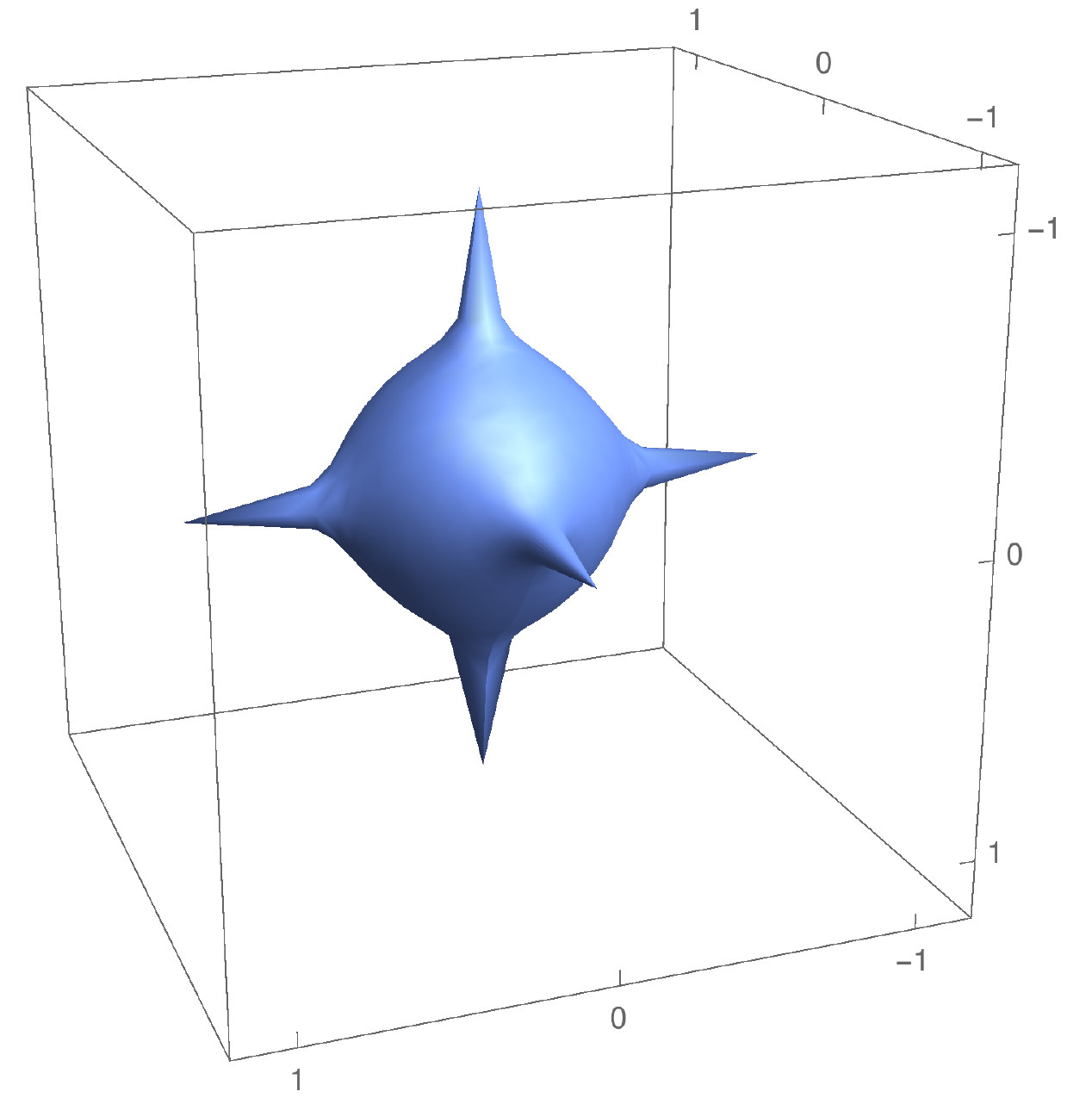} 
    \end{minipage}
    \caption{{Conformal factor $\Omega$ (left) and scalar field $\phi$ (right), for $\kappa_2=0.1$.}}
\end{figure}

\begin{figure}[t]
    \centering
        \begin{minipage}{0.3\textwidth}
        \centering
        \includegraphics[width=1\textwidth]{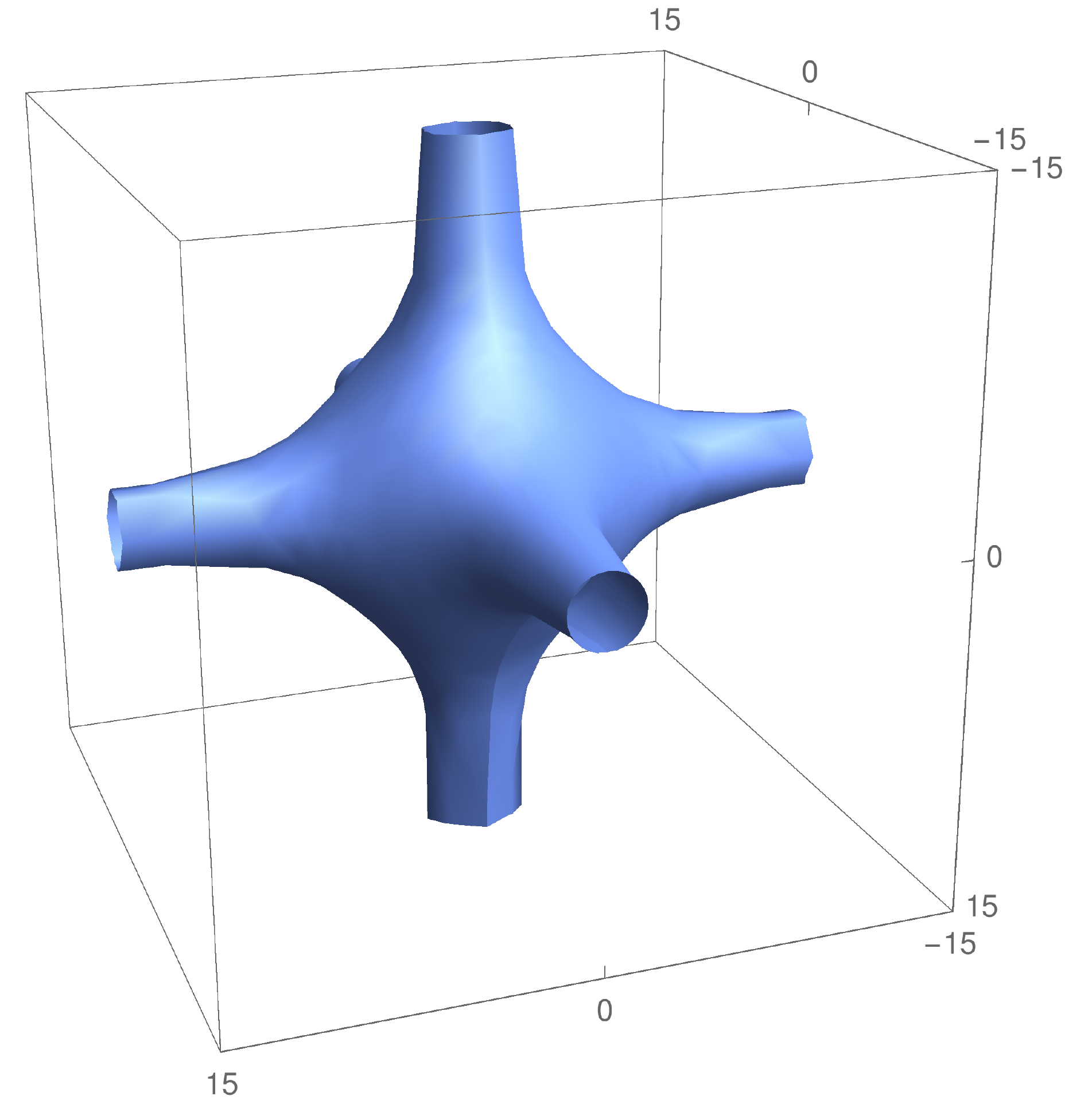} 
    \end{minipage}
    \begin{minipage}{0.3\textwidth}
        \centering
        \includegraphics[width=1\textwidth]{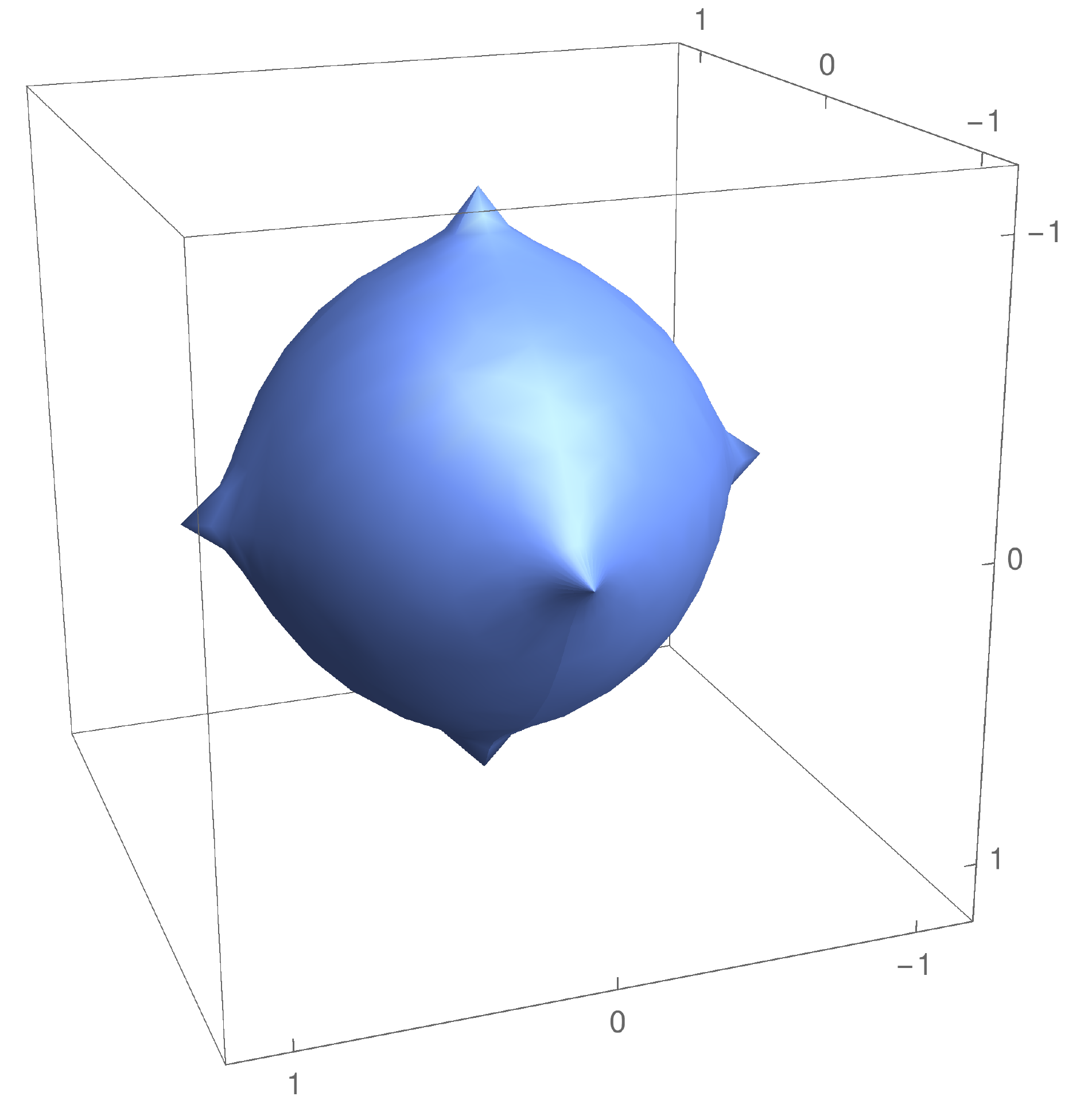} 
    \end{minipage}
    \centering
    \caption{{Conformal factor $\Omega$ (left) and scalar field $\phi$ (right), for $\kappa_2=0.4$.}}
\end{figure}

\begin{figure}[t]
    \centering
    \begin{minipage}{0.3\textwidth}
        \centering
        \includegraphics[width=1\textwidth]{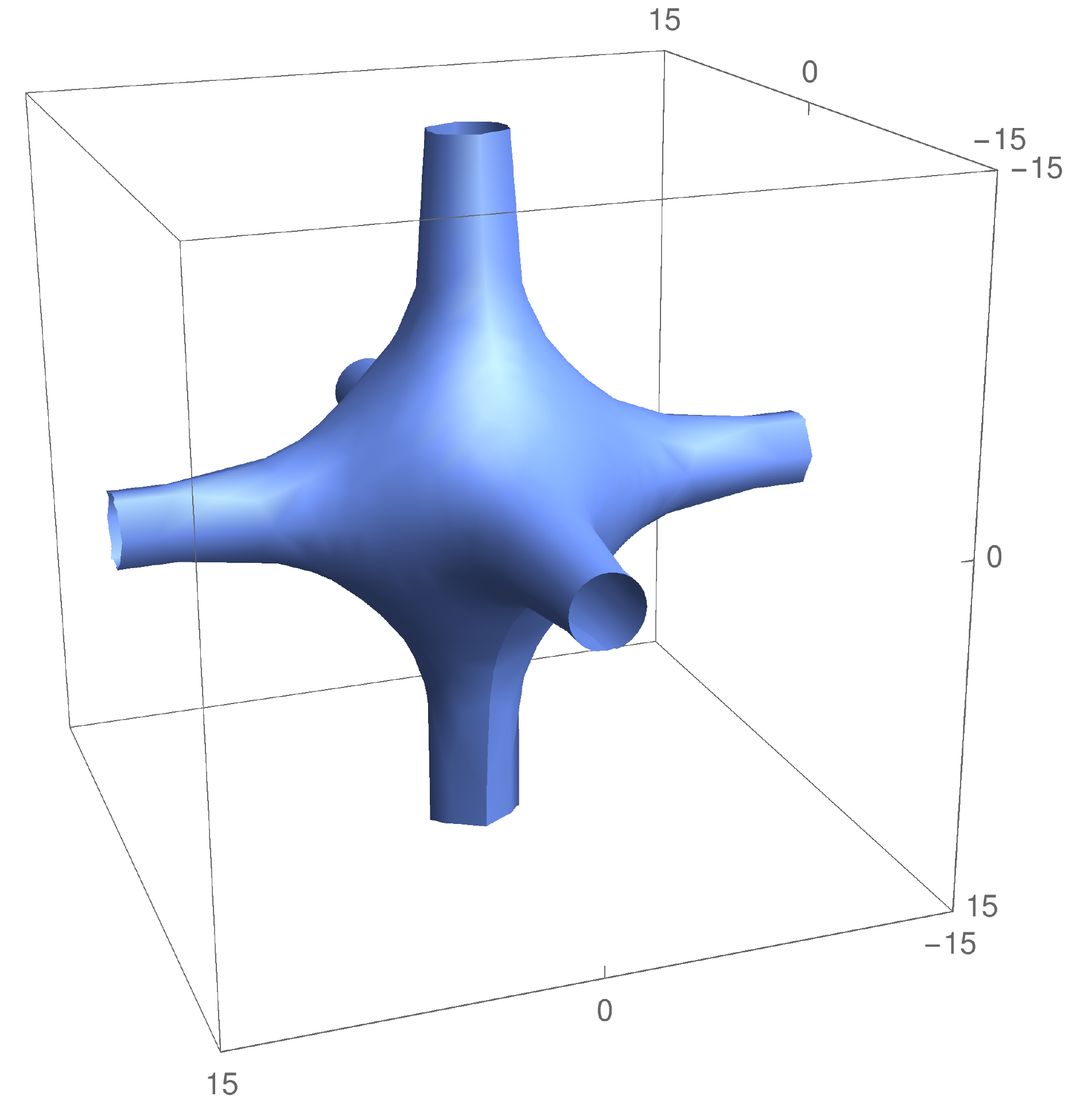} 
    \end{minipage}
    \begin{minipage}{0.3\textwidth}
        \centering
        \includegraphics[width=1\textwidth]{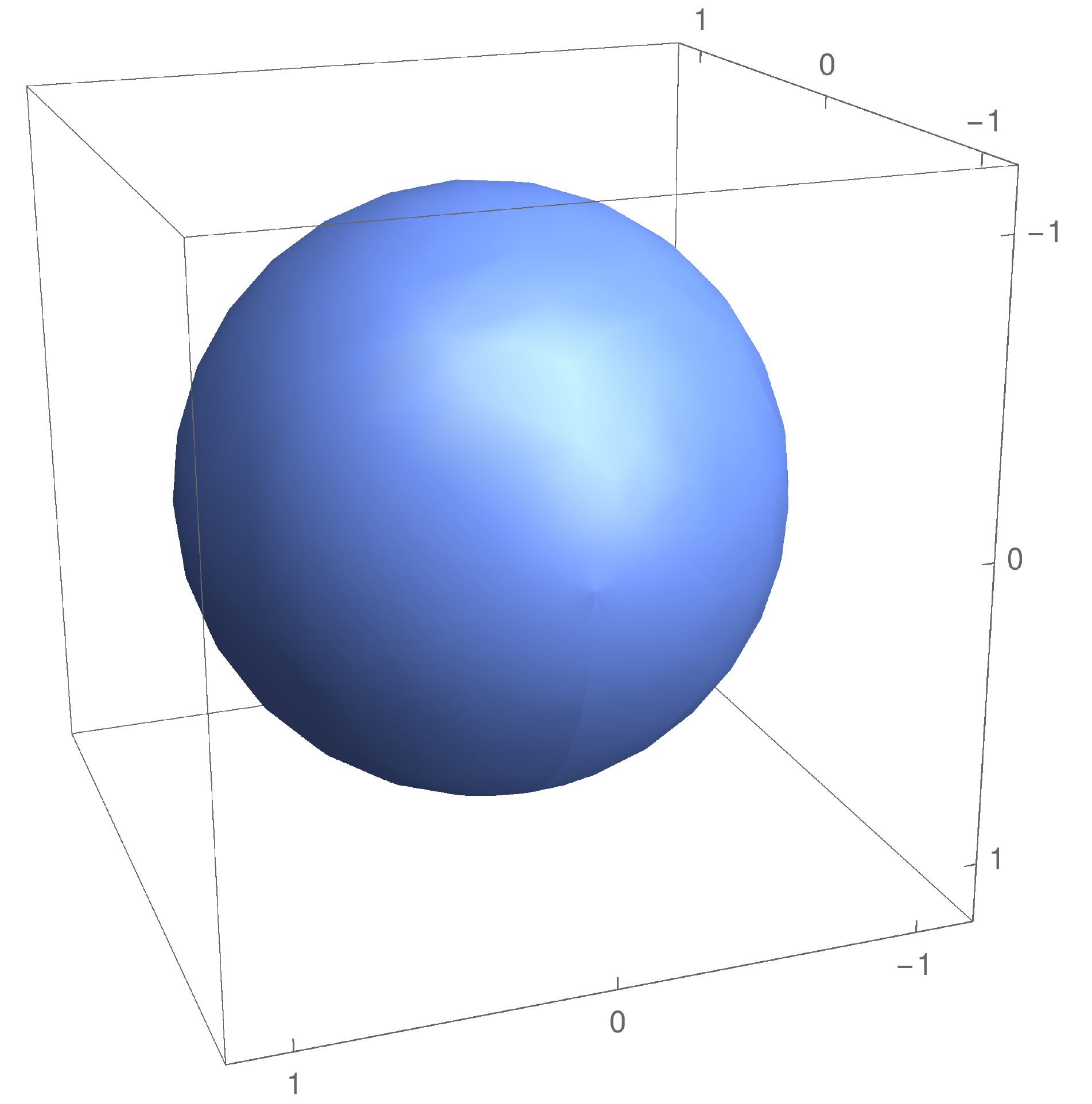}
    \end{minipage}
    \caption{{Conformal factor $\Omega$ (left) and scalar field $\phi$ (right), 
    for $\kappa_2=0.75$.}}
\end{figure}

\begin{figure}[t]
    \centering
        \begin{minipage}{0.3\textwidth}
        \centering
        \includegraphics[width=1\textwidth]{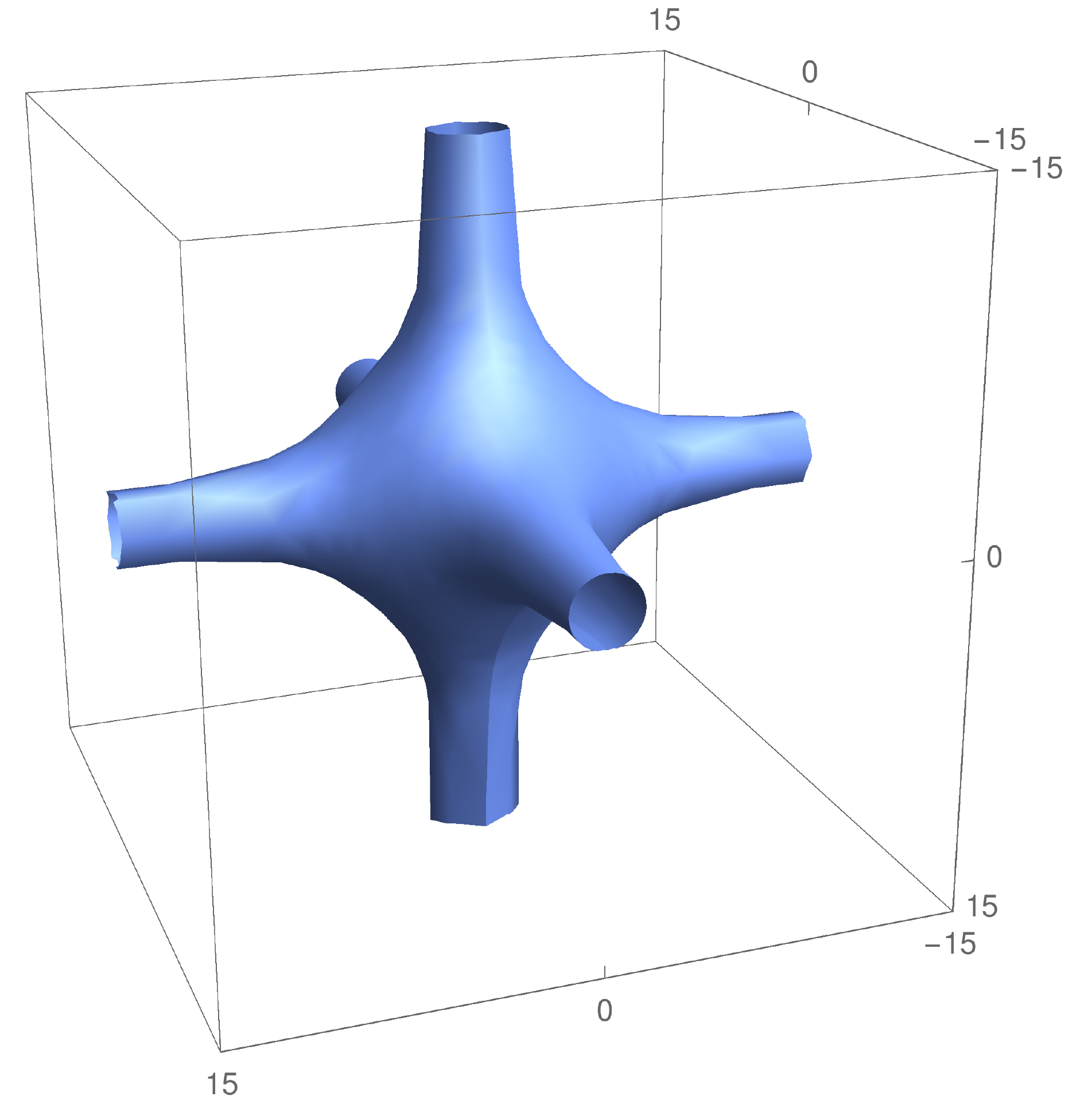} 
    \end{minipage}
    \begin{minipage}{0.3\textwidth}
        \centering
        \includegraphics[width=1\textwidth]{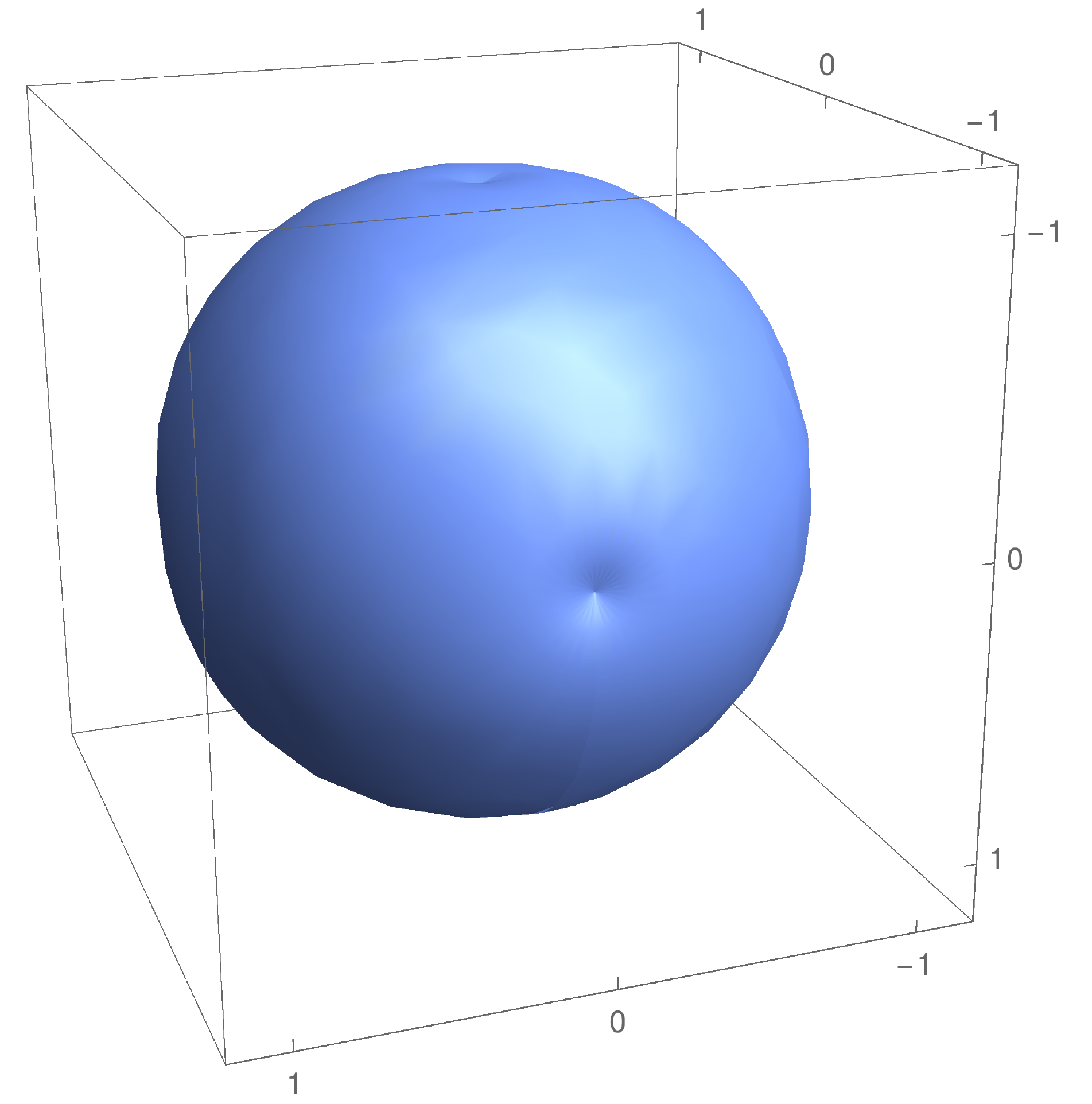} 
    \end{minipage}
    \centering
    \caption{{Conformal factor $\Omega$ (left) and scalar field $\phi$ (right), for $\kappa_2=0.9$.}}
\end{figure}

In Figure 3, for $\kappa_2 = 0.75$ there is no scalar charge on any of the masses, as the value of the scalar field is represented by a constant unit sphere. Mathematically this can be seen in Equation (\ref{eq:k1k2def}), where we have that $\kappa_1 = \kappa_2$ for $\kappa_2 = 0.75$. Setting all of the $\alpha_i$ parameters to be equal to each other, and likewise for $\gamma_i$, then implies from Equations (\ref{eq:bigai}) and (\ref{eq:bigbi}) that $\gamma_i A_i = \alpha_i B_i$. Finally it is manifest from Equation (\ref{eq:scalarcharge}) that $q_i = 0$.
For values of $\kappa_2 < 0.75$ the scalar field is largest at the positions of the masses at a maximum value of $1$, whereas for $\kappa_2 > 0.75$ the scalar field is smallest at the positions of the masses (as shown by the dimples in Figure 4). Changing the value of $\kappa_2$ can therefore be interpreted as increasing or decreasing the background value of the scalar field. We note that changing the value of $\kappa_2$ has very little effect on the geometry of the initial hypersurface itself, but that changing $\kappa_2$ or $\omega$ has a very significant effect on the scalar field distribution (with the shape of the corresponding figures again approaching a spherical shape in the limit $\omega \rightarrow \infty$). The distribution of $\phi$ in Figures 1-4 can be directly linked to the distribution of Newton's constant, $G$, via Equation (\ref{newtonG}).

\subsection{General relativistic limit}
\label{sec:bdlgrl}

We wish to investigate how (and if) the lattice cosmologies constructed above differ from their general relativistic counterparts, and how they approach them in the limit $\omega \rightarrow \infty$. Of principle interest in this regard will be the scale of the cosmological region of each of the respective solutions. In order to extract this quantity we define $a_0^{\rm BDL} \equiv \left( \chi^{2 a} \sigma^{2-2a} \right)\vert_{\rm vertex}$, where the right-hand side is being evaluated at the vertex of one of the primitive ``cells" from which the lattice is constructed (i.e. at one of the points which is furthest away from all nearby masses). A similar quantity, $a_0^{\rm GRL}$, can be constructed to measure the scale of the cosmological region in the corresponding general relativistic lattice.

\begin{figure}[t]
    \centering
        \centering
        \includegraphics[width=1\textwidth]{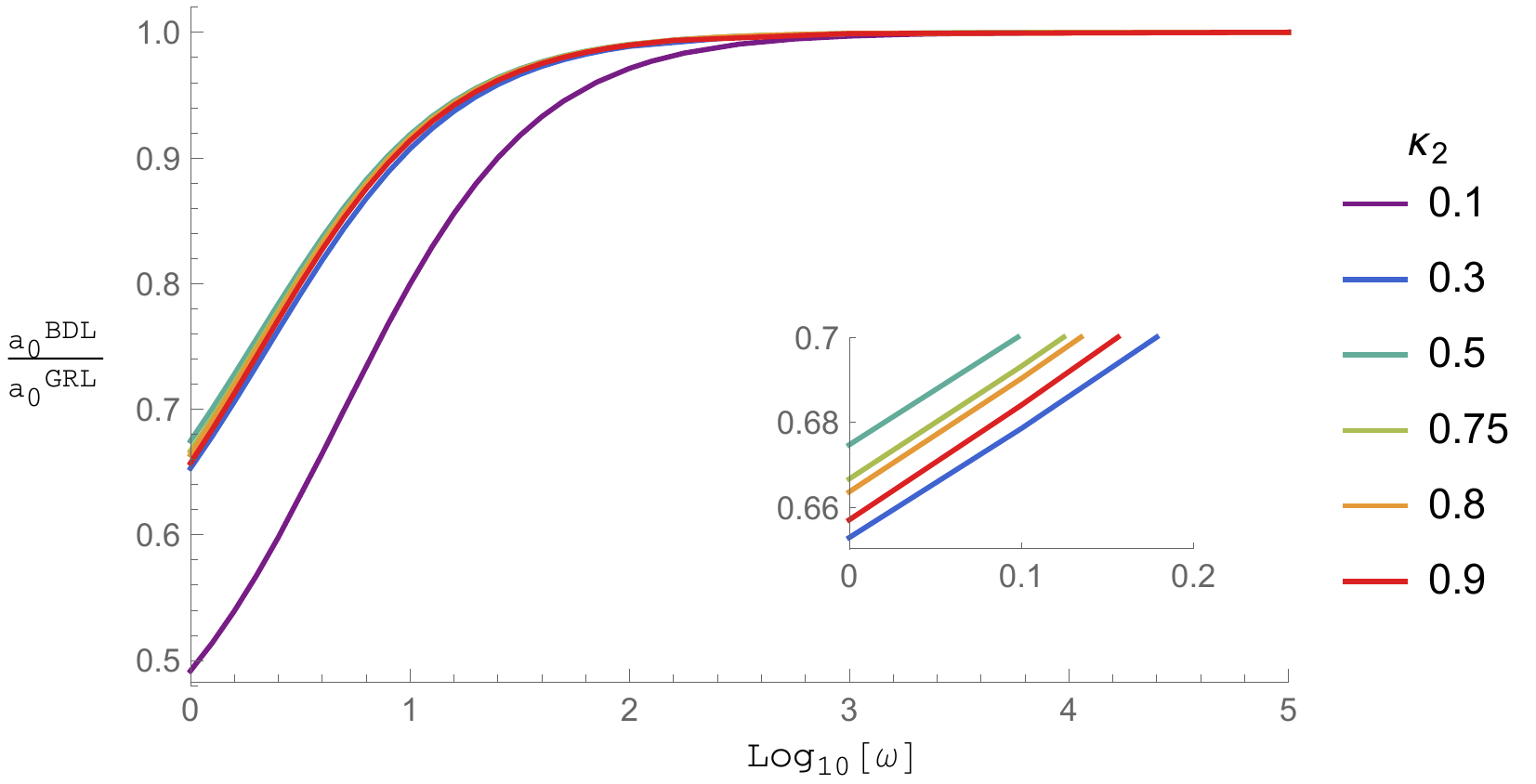} 
        \caption{{Ratio of scale factors $a_0^{\rm BDL}/a_0^{\rm GRL}$ for the BD and GR lattice cosmologies, for different values of $\omega$ and $\kappa_2$ (with $m = 9.48$ and $\gamma=1$). The inset shows a close-up of the intersection of the lines with $\kappa_2\geq 0.3$ and the y-axis.}}
        \label{fig:bdlgrl1}
\end{figure}

We now wish to compare the values of $a_0^{\rm BDL}$ and $a_0^{\rm GRL}$ for two lattices that contain the same number of objects, located at the same positions, and with the same total proper mass. We again choose to consider the 8-mass cubic lattice, as discussed in the previous section. We find that the quantity $a_0^{\rm BDL}/a_0^{\rm GRL}$ changes as a function of the coupling parameter of the theory, $\omega$, but also as a function of the parameter that controls the background value of the scalar field $\kappa_2$, where $\kappa_2 \leq 1$. In order to uniquely specify a solution in the case of the Brans-Dicke lattices we also need to specify a value for the proper mass and scalar charge of each black hole. Regarding the proper mass, we set this to be the value found in the general relativistic case, as shown in Table VI of Ref. \cite{tim1}. There, the ratio of effective mass to proper mass was found to be $0.11$. Setting the effective mass to unity, for simplicity, then yields $m_i = m = 9.48$, and here we take this to set the value of the $\alpha_i$ parameters. For the scalar charge, we find that scale factor of the cosmology is insensitive to the specific value chosen for $q_i$, therefore we can instead set $\gamma_i = 1$ for simplicity. We display our results in Figure \ref{fig:bdlgrl1}.

All our results show a convergence towards the general relativistic value of the scale factor as $\omega \rightarrow \infty$, as expected. For $\omega \lesssim 10^3$, however, our solutions are very different from the general relativistic ones, with the scale factor taking a smaller value in every case. These plots make it clear that scale of the cosmological solutions is strongly dependent on $\kappa_2$ for small values of $\omega$, but that in the limit $\omega \rightarrow \infty$ all dependence on $\kappa_2$ drops out. Finally we interpret the independence of the value of the scalar factor to the particular value of $q$ as demonstrating that the majority of the gravitational influence of each point particle is dominated by its mass, and not its scalar charge.

\section{Comparison with Brans-Dicke Friedmann cosmology}
\label{sec:flrw}

In this section we will make a comparison between the initial data for an inhomogeneous universe described above, and the corresponding homogeneous and isotropic dust-filled Friedmann cosmologies that exist in Brans-Dicke theory. Our approach to this is to compare cosmologies that contain the same total mass and background scalar field value, at a moment of time-reversal symmetry (as is implied by $K_{\mu\nu}=0$).

To do this we need to solve the field equations for homogeneous and isotropic dust-filled space-times, which are given by
\begin{eqnarray}
    H^2 &=& \frac{8\pi\rho}{3\phi}-\frac{k}{a^2}-H\frac{\dot{\phi}}{\phi}+\frac{\omega}{6}\frac{\dot{\phi}^2}{\phi^2},  \label{eq:flrw1}\\
    \frac{\ddot{\phi}}{\phi} &=& \frac{8\pi\rho}{\phi(2\omega+3)}-3H\frac{\dot{\phi}}{\phi},
    \label{eq:flrw2}
\end{eqnarray}
where $\rho = \rho_0 {a_0^3}/{a^3}$, $H = \dot{a}/a$ and where over-dots denote differentiation with respect to the proper time of comoving observers. 

We can immediately note that if we require $H=\dot{\phi}=0$ then Equation (\ref{eq:flrw1}) implies that the spatial curvature $k$ must be positive (and given by $k=8 \pi \rho a^2/3 \phi$, assuming $\rho$ and $\phi$ are both positive valued quantities). Choosing units where $k=1$, we find that there exist solutions given by \cite{flrw}
\begin{equation}
    a(t) = \frac{3 \phi_0}{8 \pi \rho_0 a_0^3} - \frac{2 \pi \rho_0 a_0^3}{\phi_0(3+2\omega)}(t-t_0)^2 \, ,
\end{equation}
where $\phi_0$ and $t_0$ are constants, and where $\phi= \phi_0 a^{-2}$. This clearly corresponds to a universe with a time-symmetric evolution, with a maximum of expansion at $t=t_0$. The intrinsic geometry of the hypersurface at maximum of expansion is therefore given by 
\begin{equation} \label{dshom}
    ds^2 = \frac{9 \phi_0^2}{64 \pi^2 \rho_0^2 a_0^6} d\bar{s}^2
    = \frac{9 \pi^2 \phi_0^2}{16 M^2} d\bar{s}^2
    \, ,
\end{equation}
where in the last equality we have used the fact that $\rho = {M}/{V} = {M}/{2\pi^2 a^3}$, and where $M$ and $V$ are the mass and spatial volume of the hypersphere. This gives us the scale of the maximum of expansion of such a universe in terms of the total mass of the matter content, $M$, and the constant associated with the scalar field, $\phi_0$.

In order to find suitable inhomogeneous solutions to compare to Equation (\ref{dshom}) we choose to consider solutions in which $M=N m$, where $N$ is the total number of identical point-like masses in the inhomogeneous solution and $m$ is the proper mass of each of them. This condition means we compare cosmological models that contain the same total mass. The second condition we need to implement is on the value of $\phi_0$. To do this, we require that the background value of the scalar field in the inhomogeneous solutions must equal that of the Friedmann cosmology. Using the fact that the scalar field in the inhomogeneous solutions is given by $\phi=\chi^s \sigma^{-s}$, and equating it to the value of $\phi$ at the maximum of expansion of the Friedmann models, we find
\begin{equation}
    \chi^s \sigma^{-s} = \frac{16 M^2}{9 \pi^2 \phi_0} \qquad \Rightarrow \qquad \phi_0 = \frac{16 M^2}{9 \pi^2 \chi^s \sigma^{-s}} \, ,
\end{equation}
where $\chi$ and $\sigma$ are to be given values associated with the cosmological background. There is clearly some freedom in choosing how this should be done, as both quantities are in general non-constant functions of spatial position. Here we proceed as in the previous section and choose to take their value at the location that is farthest from all masses, at the vertex of one of the primitive cells of the lattice, as this is the closest thing to taking a ``background value'' in an inhomogeneous cosmology. Correspondingly, we will also evaluate the scale factor in the inhomogeneous solutions at the same point, in order to make a fair comparison. This means that we can now write the scale factors for Brans-Dicke lattice (BDL) cosmologies, and the corresponding Brans-Dicke Friedmann (BDF) cosmologies, as
\begin{equation}
    a_0^{\rm BDL} = \chi^{2a} \sigma^{2-2a} \qquad {\rm and} \qquad a_0^{\rm BDF} = \frac{4 N m}{3\pi \chi^s \sigma^{-s}} \, ,
\end{equation}
where $\chi$ and $\sigma$ are both to be evaluated at the locations farthest from all masses. A comparison of $a_0^{\rm BDL}$ and $a_0^{\rm BDF}$ will then give a numerical quantification for the effects of structurisation of matter in Brans-Dicke cosmologies.

\begin{figure}[t]
    \centering
        \centering
        \includegraphics[width=1\textwidth]{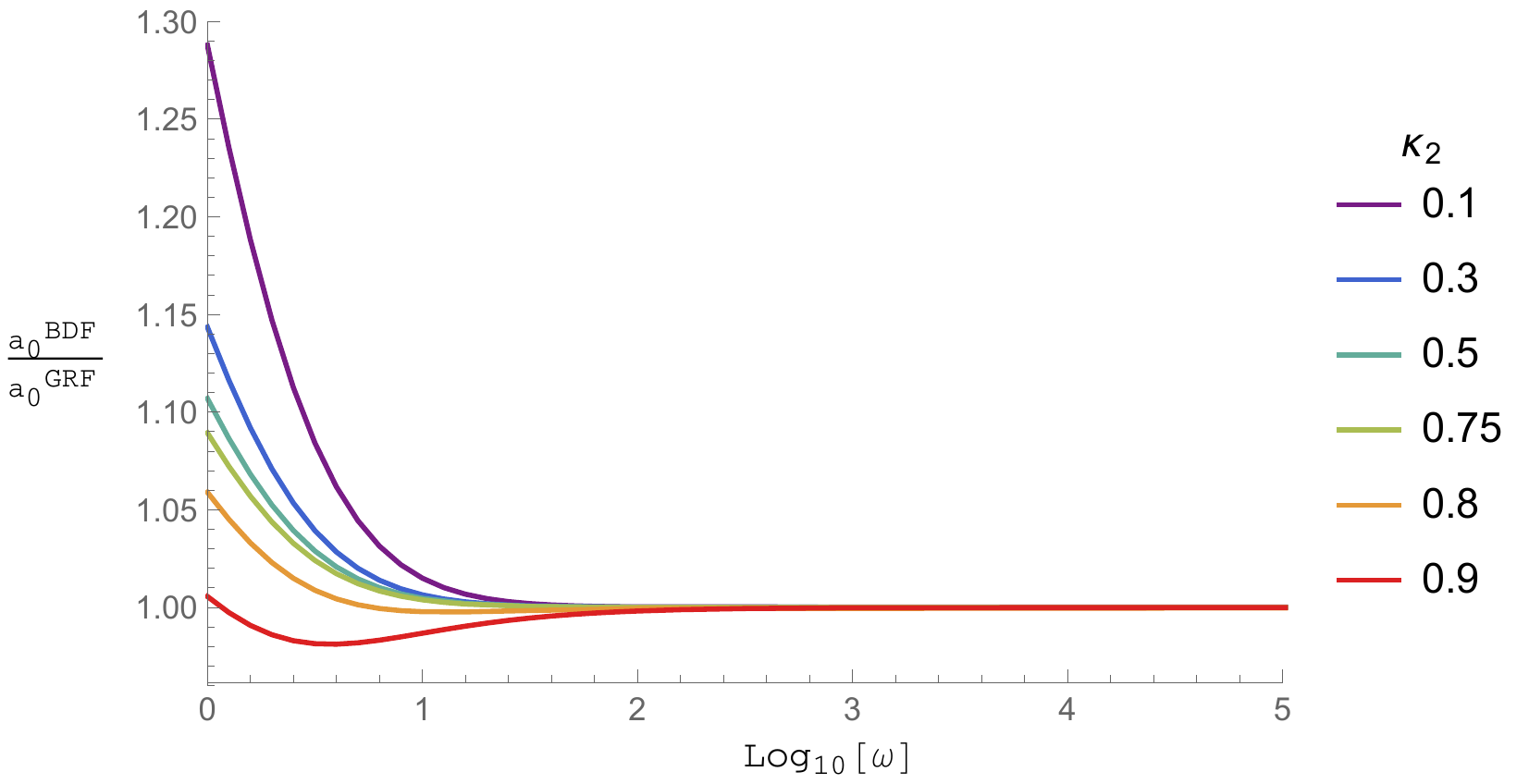}
        \caption{{Ratio of scale factors $a_0^{\rm BDF}/a_0^{\rm GRF}$ for the BD and GR Friedmann cosmologies, for different values of $\omega$ and $\kappa_2$ (with $m = 9.48$ and $\gamma=1$).}}
        \label{fig:bdfgrf}
\end{figure}

In order to consider specific models, we again choose to consider the 8-mass model discussed above, and again choose the proper mass of each of our sources to have $m_i=m=9.48$ (so that the total mass in the corresponding Friedmann solution is $M=75.84$). We also set the parameter $\gamma_i= 1$ for each particle, as before. Under these conditions, we plot two quantities. The first is the ratio of scales in the Friedmann cosmologies for the Brans-Dicke theory and general relativistic case, in Figure \ref{fig:bdfgrf}. The value of $a_0^{\rm BDF}$ approaches the general relativistic value as $\omega \rightarrow \infty$, as expected, and similarly to Figure \ref{fig:bdlgrl1}. For small $\omega$ the scale of the Brans-Dicke Friedmannian cosmology is much larger than its general relativistic counterpart, which contrasts the behaviour in Figure \ref{fig:bdlgrl1}.

\begin{figure}[t]
    \centering
        \includegraphics[width=1\textwidth]{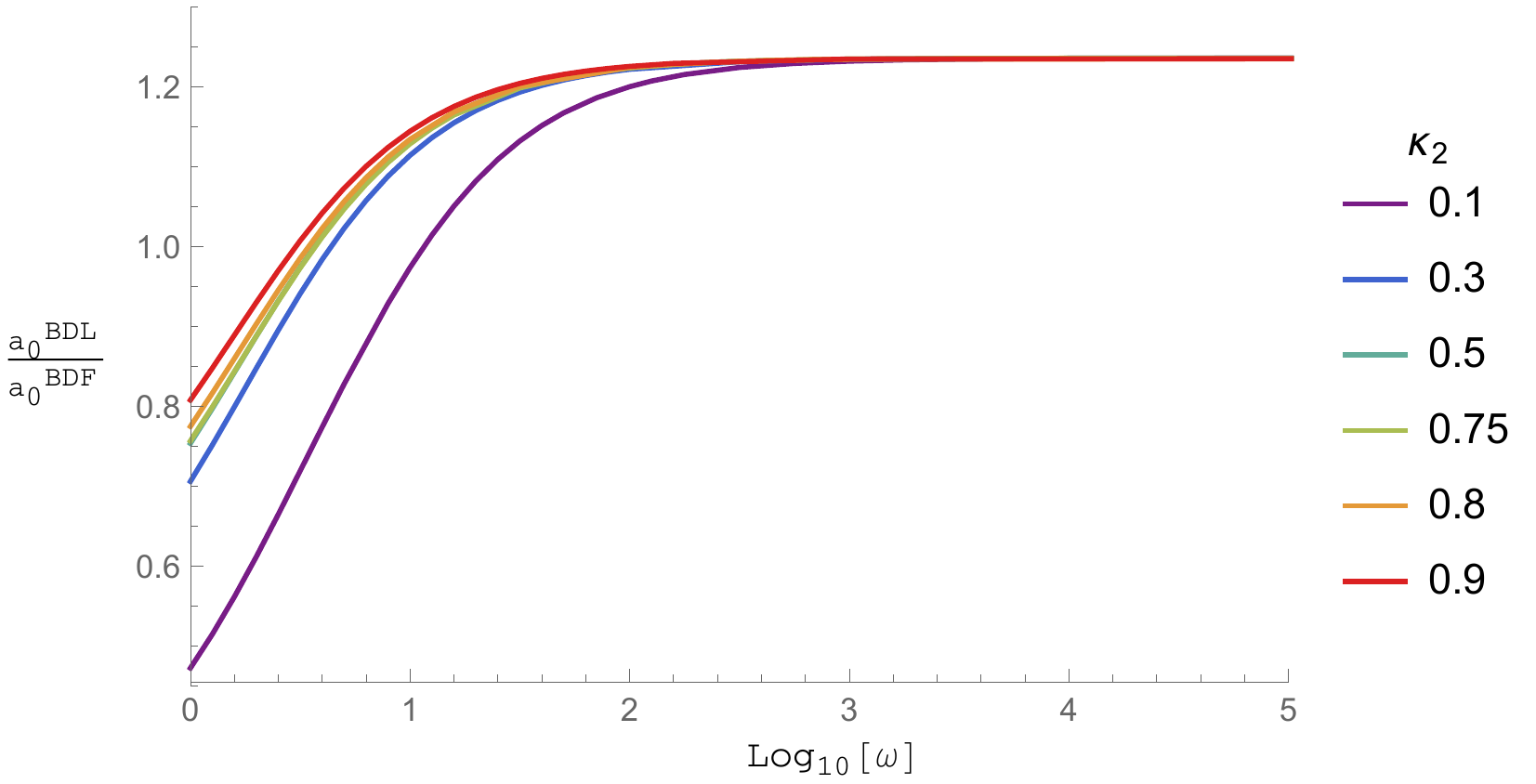}
        \caption{{The ratio of scales $a_0^{\rm BDL}/a_0^{\rm BDF}$, for the Brans-Dicke lattice and Friedmann cosmologies (with $m = 9.48$ and $\gamma = 1$).
       }}
        \label{fig:bdlbdf}
\end{figure}

The second quantity is the ratio of scales in the lattice and Friedmann cosmologies in just the Brans-Dicke theory, in Figure \ref{fig:bdlbdf}. The value of $a_0^{\rm BDL}/a_0^{\rm BDF}$ for the 8-mass lattice can be clearly seen to approach the general relativistic value of $1.236$ \cite{tim1} as $\omega \rightarrow \infty$. For small $\omega$, on the other hand, the scale of the lattice cosmology is much smaller than its Friedmann counterpart, by as much as $50 \%$ for $\kappa_2 = 0.1$. It is interesting that there are small values of $\omega$ where the theory is far from a general relativistic one (for example, $\omega = 10, \kappa_2 = 0.1$) but the process of constructing either a lattice cosmology or a fluid one makes no difference as far as the ratio of scale factors is concerned (for these values this ratio is approximately 1). We can therefore construct cosmologies where there is no backreaction. However, the reader should also note that the Brans-Dicke coupling parameter is constrained to be $\omega \gtrsim$40,000 to $2\sigma$, from solar system tests \cite{solsystests}. Our global scale of our models show rapid convergence to their general relativistic counterparts for values of $\omega$ this large, and should therefore should not be expected to give any detectable difference on very large scales if the governing theory is to be compatible with solar system constraints. Nevertheless, in such cases the scalar field can still vary considerably in the vicinity of the masses themselves, and may also give potential deviations from general relativity in their future evolution, as more extreme environments are encountered. Theory independent variations on the Newton's constant can also be used to constrain these models, and can be found in Ref. \cite{uzan1}-\cite{uzan2}. Such constraints tend to be imposed on the time variation of $G$, and are found from a number of different observations to be constrained at the level $\dot{G}/G \lesssim 10^{-12}$ per year. Numerical evolution of our initial data would allow us to investigate the behaviours that are compatible with these bounds, but this will be left for future studies.

\section{Discussion}
\label{sec:discussion}

We have provided, for the first time, exact initial data for a cosmological model in scalar-tensor theories of gravity that contains a regular array of point-like particles. This was achieved by first deriving the relevant constraint equations (in Section \ref{sec:constraints}), and then by imposing the condition that the extrinsic curvature vanishes on the initial hypersurface. We found a simple set of solutions to these constraint equations, in terms of a pair of conformal factors, which reduces in the appropriate limits to the known static, spherically symmetric vacuum Brans solution. Comparison to this exact solution then allowed us to derive expressions for the proper mass and scalar charge for each of the particles in our cosmologies. We find that the scalar charge of each of the black holes vanishes in the general relativistic limit, when $\omega \to \infty$, and that the spatial variation of Newton's constant depends on both the scalar charge of the individual bodies as well as a cosmological background value.

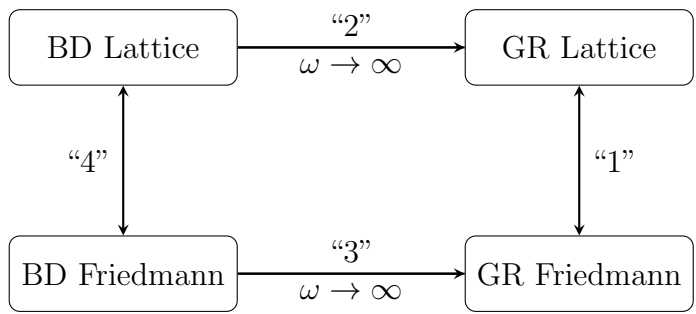
\begin{figure}[h!] 
\begin{center}

\begin{tikzpicture}[node distance=2cm, every node/.style={anchor=center}]

\node (bdl) [startstop] {BD Lattice};

\node (grl) [startstop, right of=bdl,xshift=4cm] {GR Lattice};

\node (bdf) [startstop, below of=bdl,yshift=-1cm] {BD Friedmann};

\node (grf) [startstop, right of=bdf,xshift=4cm] {GR Friedmann};

\draw [arrow1] (bdl) -- node[anchor=north] {$\omega \to \infty$} (grl);
\draw [line] (bdl) -- node[anchor=south] {``2"} (grl);
\draw [arrow2] (bdl) -- node[anchor=east] {``4"} (bdf);
\draw [arrow1] (bdf) -- node[anchor=north] {$\omega \to \infty$} (grf);
\draw [line] (bdf) -- node[anchor=south] {``3"} (grf);
\draw [arrow2] (grl) -- node[anchor=west] {``1"} (grf);

\end{tikzpicture}
\end{center}
\caption{Schematic diagram showing four different cosmologies, and the comparisons that are possible between them.} \label{fig:diagram9}
\end{figure}

We have considered the general relativistic limit of a specific realisation of our lattice solution (Figure \ref{fig:bdlgrl1}), as well as a comparison between the Friedmann solutions of Brans-Dicke and general relativity at a maximum of expansion (Figure \ref{fig:bdfgrf}), and a comparison between discrete and continuous cosmological solutions in Brans-Dicke theory alone (Figure \ref{fig:bdlbdf}). This was done in Sections \ref{sec:constraints} and \ref{sec:flrw}. In all cases it was found that our new solutions approach the expected general relativistic limits as $\omega \rightarrow \infty$, and that order one deviations from the general relativistic results were possible when $\omega$ was small. Our solutions were also found to be sensitive to the value of the parameter $\kappa_2$, which controls the background value of the scalar field. These results can be considered as three comparisons between a set of four cosmological models, as shown in Figure \ref{fig:diagram9}.

The branch labelled ${``1"}$ corresponds to the comparison between discrete and continuous cosmologies, as initiated in Ref. \cite{tim1} and reviewed in Ref. \cite{durkreview}. The work in Sections \ref{sec:constraints} and \ref{sec:flrw} of this paper provide the first (and currently only) steps to understand branches ${``2"}$, ${``3"}$ and ${``4"}$ of this graph. Our analyses were performed by calculating the ratio of the line-elements in the respective models and theories, at the vertex of a primitive cell of the lattice. Future steps to understanding this problem further would be to investigate how the number and distribution of massive bodies affects the cosmological properties of the space-time in these theories, and to numerically evolve this initial data to recover the geometry of the full space-time. We leave this for subsequent studies.

\section*{Acknowledgements}

JD and TC both acknowledge support from the STFC under grant
STFC ST/N504257/1.

\end{document}